\documentclass[11pt]{article}
\usepackage[total={6.5in,8.75in}, top=1in, left=0.8in, right=0.8in, bottom=1in, includefoot]{geometry}
\usepackage{titlesec}
\usepackage{amsmath}
\usepackage{amssymb}
\usepackage{authblk}
\usepackage{graphicx}
\usepackage{bm}
\usepackage[numbers,sort&compress]{natbib}
\usepackage{caption}
\titleformat{name=\section}
{\normalfont\Large\bfseries}
{\thesection}
{1em}
{}

\titleformat{name=\subsection}
{\normalfont\bfseries}
{\thesubsection}
{1em}
{}
\titlespacing{\subsection}{0pt}{*0}{0pt}

\newcommand{\alternative}[1]{}

\usepackage[numbers]{natbib}
\usepackage[%
colorlinks,bookmarksopen,bookmarksnumbered,citecolor=blue,urlcolor=blue
]{hyperref}

\title{A practical guide to methodological considerations in the controllability of structural brain networks}
\author[a,b]{Teresa M. Karrer}
\author[b,*]{Jason Z. Kim}
\author[c,*]{Jennifer Stiso}
\author[c]{Ari E. Kahn}
\author[d]{Fabio Pasqualetti}
\author[a,e,f]{Ute Habel}
\author[b,g,h,i,j,k,l]{Danielle S. Bassett}
\affil[a]{Department of Psychiatry, Psychotherapy and Psychosomatics, Faculty of Medicine, RWTH Aachen, Germany.}
\affil[b]{Department of Bioengineering, School of Engineering \& Applied Science, University of Pennsylvania, Philadelphia, PA 19104, USA.}
\affil[c]{Department of Neuroscience, Perelman School of Medicine, University of Pennsylvania, Philadelphia, PA 19104, USA.}
\affil[d]{Department of Mechanical Engineering, University of California, Riverside, CA 92521, USA.}
\affil[e]{JARA - Translational Brain Medicine, Aachen, Germany.}
\affil[f]{Institute of Neuroscience and Medicine: JARA-Institute Brain Structure Function Relationship (INM 10), Research Center Jülich, Jülich, Germany.}
\affil[g]{Department of Physics and Astronomy, College of Arts \& Sciences, University of Pennsylvania, Philadelphia, PA 19104, USA.}
\affil[h]{Department of Neurology, Perelman School of Medicine, University of Pennsylvania, Philadelphia, PA 19104, USA.}
\affil[i]{Department of Psychiatry, Perelman School of Medicine, University of Pennsylvania, Philadelphia, PA 19104, USA.}
\affil[j]{Department of Electrical and Systems Engineering, School of Engineering \& Applied Science, University of Pennsylvania, Philadelphia, PA 19104, USA.}
\affil[k]{Santa Fe Institute, Santa Fe, NM 87501, USA.}
\affil[l]{To whom correspondence should be addressed: dsb@seas.upenn.edu}
\affil[*]{These two authors contributed equally.}
\date{}

\begin{document}
\maketitle
\bibliographystyle{unsrtnat}

\begin{abstract}
Predicting how the brain can be driven to specific states by means of internal or external control requires a fundamental understanding of the relationship between neural connectivity and activity. Network control theory is a powerful tool from the physical and engineering sciences that can provide insights regarding that relationship; it formalizes the study of how the dynamics of a complex system can arise from its underlying structure of interconnected units. Given the recent use of network control theory in neuroscience, it is now timely to offer a practical guide to methodological considerations in the controllability of structural brain networks. Here we provide a systematic overview of the framework, examine the impact of modeling choices on frequently studied control metrics, and suggest potentially useful theoretical extensions. We ground our discussions, numerical demonstrations, and theoretical advances in a dataset of high-resolution diffusion imaging with 730 diffusion directions acquired over approximately 1 hour of scanning from ten healthy young adults. Following a didactic introduction of the theory, we probe how a selection of modeling choices affects four common statistics: average controllability, modal controllability, minimum control energy, and optimal control energy. Next, we extend the current state of the art in two ways: first, by developing an alternative measure of structural connectivity that accounts for radial propagation of activity through abutting tissue, and second, by defining a complementary metric quantifying the complexity of the energy landscape of a system. We close with specific modeling recommendations and a discussion of methodological constraints. Our hope is that this accessible account will inspire the neuroimaging community to more fully exploit the potential of network control theory in tackling pressing questions in cognitive, developmental, and clinical neuroscience.
\end{abstract}


\newpage

\section{Introduction}
The brain is a complex system of interconnected units that dynamically transitions through diverse activation states supporting cognitive function \cite{deco2011emerging}. Understanding the mechanisms and processes that give rise to these trajectories through state space is crucial for intervening in disease to restore cognitive functioning \cite{deco2014great}. One relevant factor enabling such rich neural dynamics is the network architecture of the underlying structural substrate \cite{hermundstad2011learning, hermundstad2013structural, hermundstad2014structurally}. Yet, the exact mechanisms by which the physical architecture of the brain both supports and constrains its function remain largely unknown \cite{rajan2016recurrent, levy1999distributed, fiete2010spike}. 
\\
\\
Recent advances in network control theory offer a formal means to study how the temporal dynamics of a complex system emerges from its underlying network structure \cite{kailath1980linear, liu2011controllability,pasqualetti2014controllability}. Applying this theory to the brain requires that one first builds a network model in which brain regions (nodes) are anatomically connected to one another (edges) \cite{bullmore2009complex,bassett2018nature}. The state of the brain network system is then reflected in the pattern of neurophysiological activity across network nodes, and state trajectories represent the temporal sequence of brain states that the system traverses \cite{shenoy2011dynamical, freeman1994characterization}. With definitions of the network and its state in hand, we can consider the problem of network controllability, which in essence amounts to asking how the system can be driven to specific target states by means of internal or external control input \cite{kalman155070controllability}. In the context of the brain, such input can intuitively take the form of electrical stimulation \cite{muldoon2016stimulation,stiso2018white,solomon2018medial,medaglia2018network,khambhati2019functional}, task modulation \cite{cui2018optimization,muldoon2018locally,cornblath2019sex}, or other perturbations from the world or from different portions of the body \cite{dum2016motor,goyal2015feeding}. Practically, network control theory and its associated toolkit enables us to study the general role of brain regions in controlling neural dynamics in diverse scales and species \cite{yan2017network, towlson2018caenorhabditis, wiles2017autaptic,kim2018role}, and in both health \cite{gu2015controllability,shine2019human} and disease \cite{jeganathan2018fronto,bernhardt2019temporal} or injury \cite{gu2017optimal}. Moreover, the approach can be used to determine the patterns of input required to induce specific state transitions necessary for behavior \cite{gu2017optimal,betzel2016optimally,cui2018optimization,stiso2018white}.
\\
\\
Network control theory offers three primary advantages over traditional approaches to the study of brain network function. First, the multi-modal nature of the theoretical framework explicitly enforces a simultaneous study of brain structure and function, in contrast to approaches that characterize each separately and then assess statistical covariance. Second, network control theory exceeds the often purely descriptive approach of network science \cite{bassett2009human, medaglia2015cognitive, van2016comparative} by building a generative model parameterized by both a network's spatial features and its temporal features \cite{braun2018maps}. The model then offers predictions of the brain's response to both endogenous and exogenous input signals. In the case of the former, the model could hypothetically prove useful in understanding how the brain enacts cognitive control to reach task-relevant cognitive states \cite{gu2015controllability,cui2018optimization,cornblath2019sex}. In the context of the latter, the model could similarly prove useful in informing neuromodulation for the treatment of neurological and psychiatric disorders \cite{braun2018maps}. Third, initial studies applying network control theory in neuroscience demonstrate that network controllability is a useful marker of brain dynamics, from quantifying the capacity of different brain regions to alter whole-brain dynamics \cite{gu2015controllability}, over demonstrating that this capacity grows with development \cite{tang2017developmental}, to finding that controllability is linked to executive functioning \cite{cui2018optimization,cornblath2019sex,braun2019brain}. Moreover, applications of the theory to data collected during invasive neuromodulation regimens demonstrate the utility of the theory in predicting response to electrical stimulation in practice \cite{stiso2018white,khambhati2019functional} and in theory \cite{muldoon2016stimulation}.
\\
\\
In light of the promising applicability of network control theory in neuroscience, we wish to provide a systematic overview of how the framework can be used to study the controllability of neural dynamics. This primer is constructed so as to offer neuroscientists some basic intuitions regarding the foundational concepts, and to guide them through the necessary prerequisites and considerations. For a more technical introduction that nevertheless remains heavily motivated by neuroscience, we refer the interested reader to \cite{tang2018colloquium, kim2019linear}; and for further information about the underlying mathematics (which remains agnostic to the application domain), we refer the reader to \cite{kailath1980linear,pasqualetti2014controllability}. Because the application of network control theory can be formulated in several ways, we systematically probe how diverging theoretical assumptions and possible modeling choices influence controllability metrics and the estimated energy of state transitions. For example, we consider discrete and continuous time systems, methods for system stabilization, the time horizon for control, and the set of control nodes. We complement these studies with specific recommendations for best practices, which depend in no small part upon the nature of the neuroscientific question being investigated. To further stimulate research in this exciting field, we suggest a few useful extensions of the theoretical framework, such as alternative estimates of structural connectivity and a complementary metric that quantifies the complexity of the energy landscape. 

\section{Theoretical framework}

\subsection{Network control theory} 
\begin{figure*} [hbt!]
	\centering
	\includegraphics[width=0.95\textwidth]{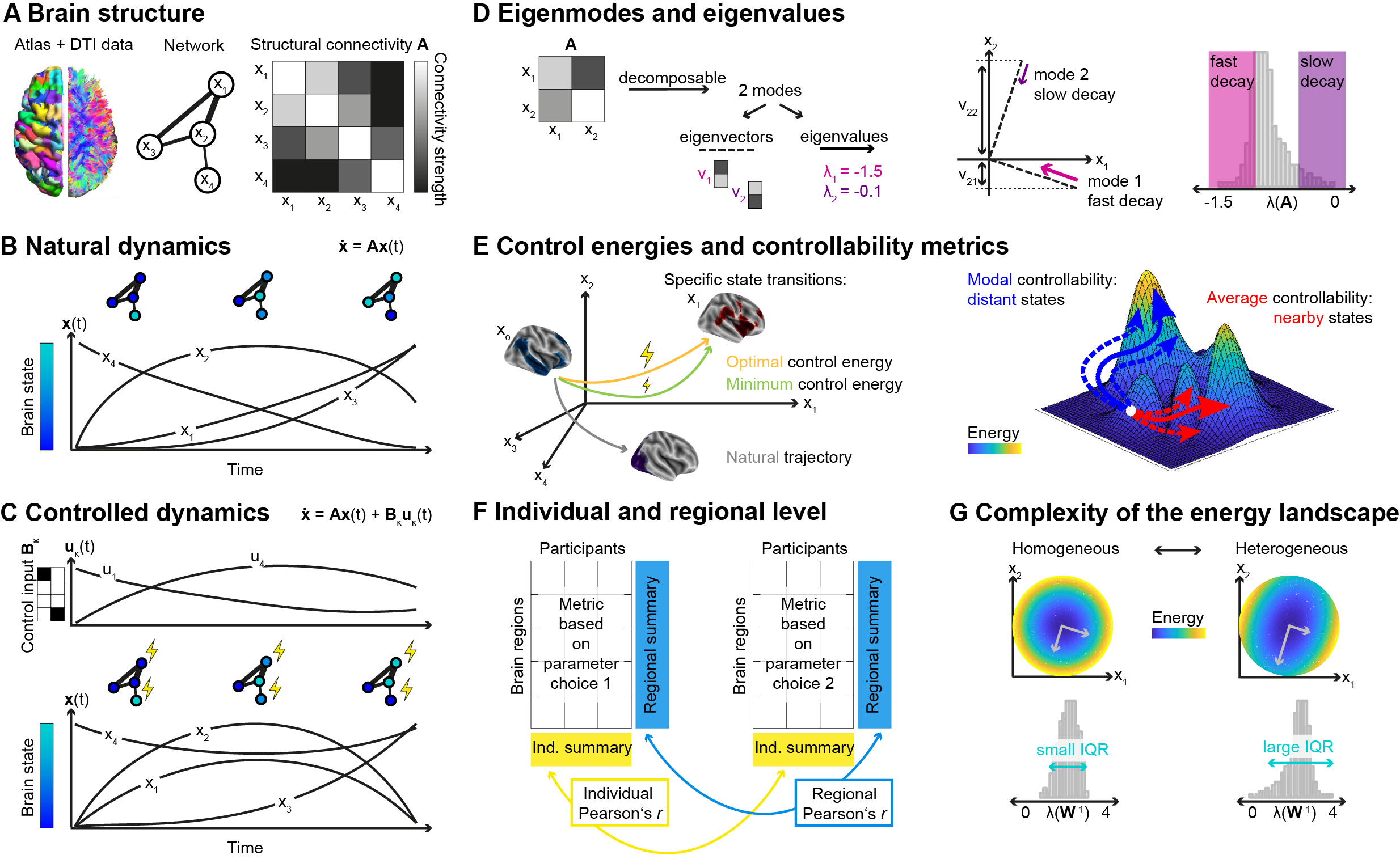}
	\caption{\textbf{Schematics of network control theory and relevant concepts.}
		\emph{(A)} \emph{Structural brain network construction.} Brain atlas and diffusion imaging data define the nodes and edges of the structural connectivity matrix $\bm{A}$. \emph{(B)} \emph{Natural dynamics of the brain.} The temporal evolution of brain states, such as the magnitude of neurophysiological activity across brain regions, is modeled as a linear function of brain structure. The area under the curve illustrates the impulse response and thus, average controllability of brain region \#4. \emph{(C)} \emph{Controlled dynamics.} The state trajectory additionally depends on control input injected into the system. The control input matrix $\bm{B}_\kappa$ determines the nodes into which a control signal $\bm{u}_\kappa$ is injected (yellow flash) over time. The area under the curve of the control energy signals corresponds to the control energy required by the given state transition. \emph{(D)} \emph{Activity modes of a system.} The structural connectivity matrix $\bm{A}$ can be decomposed into $N$ eigenvectors and eigenvalues that determine the system's dynamics. Eigenvectors determine the supported modes of activity; eigenvalues determine the rate of decline of their associated mode. Brain region $i$'s controllability $v_{i,j}$ of mode $j$ corresponds to a projection of the $j$th eigenvector onto the dimension spanned by brain region $i$. \emph{(E)} \emph{Control energies and controllability metrics.} \emph{(Left)} Control energy for specific state transitions. Here we illustrate the minimum control energy required to drive the brain from a specific initial state to a specific target state using a particular control node set. The optimal control energy additionally constrains the size of the state trajectory. \emph{(Right)} Control strategies potentially examining all possible state transitions (dashed arrows). Average controllability has been previously described as a brain region's ability to control nearby states that require little energy. Modal controllability has been previously described as a brain region's ability to control distant states that require more energy. \emph{(F)} \emph{Controllability metrics and control energies can be relevant on an individual and regional level.} To examine both levels separately, we will summarize statistics across brain regions and individuals, respectively. For consistency between two parameter choices such as discrete- and continuous-time systems, we will calculate the Pearson correlation between individual (regional) values extracted from one parameter choice and those extracted from the second parameter choice. \emph{G)} \emph{Complexity of the energy landscape.} The landscape of possible minimum control energy trajectories is determined by the eigenvalues of the inverse of the controllability Gramian $W_{\kappa, T}$. We used the variability of the eigenvalues to quantify the heterogeneity of the energy landscape. Abbreviations: IQR, interquartile range.}
	\label{fig:fig1}
\end{figure*}

The core of the theoretical framework is the structural network of neurons (or larger neural units) in the brain that allows the activity of a brain region to diffuse and change the activity of connected brain regions (Fig. \ref{fig:fig1}A). Here, we introduce a mathematical model that describes the natural dynamics of a complex linear system (Fig. \ref{fig:fig1}B). Formally, the temporal evolution of network activity is modeled as a linear function of its connectivity:
\begin{equation}\label{dynamics}
\bm{\dot{x}} = \bm{Ax}(t),
\end{equation}
where ${\bm{x}(t)}$ is a vector of size ${N\times 1}$ that represents the state of the system. Here we operationalize the system's state to reflect the magnitude of the neurophysiological activity of the ${N}$ brain regions at a single point in time. Over time, $\bm{x}(t)$ denotes the state trajectory, which is the temporal sequence of states or activity patterns that is traversed by the system. The adjacency matrix $\bm{A}$ is of size ${N\times N}$, and denotes the relationships between the system elements. Here, we operationalize that relation as the structural connectivity between each pair of brain regions.
\\
\\
Next, we extend this model to account for controlled dynamics, which occur when the brain is induced to deviate from its natural trajectory by the injection of internal or external input signals (Fig. \ref{fig:fig1}C). In this case, the temporal dynamics of a system additionally depends on the control energy injected into a set of nodes across time
\begin{equation}\label{key}
\bm{\dot{x}} = \bm{Ax}(t)+\bm{B}_{\kappa}\bm{u}_{\kappa}(t).
\end{equation}
Here, $\bm{B}_{\kappa}$ is a matrix of size ${N\times m}$ that denotes the set of $m$ control nodes or brain regions into which we wish to inject inputs. This matrix consists of $m$ indicator vectors, each having set only the ${i}$-th element to 1, corresponding to a control node. If we control all brain regions, $\bm{B}_{\kappa}$ corresponds to the $N \times N$ identity matrix with ones on the diagonal and zeros elsewhere. If we control only a single brain region $i$, $\bm{B}_{\kappa}$ reduces to a single $N \times 1$ vector with a one in the $i$-th element and zeros elsewhere. The term ${\bm{u}_{\kappa}(t)}$ is a vector of $m$ functions of size ${m \times 1}$ denoting the control input, which is the amount of input injected into each of the $m$ control nodes at each time point $t$. Over time, $\bm{u}_{\kappa}(t)$ denotes the injected control input over time.
\\
\\
For the interested reader, we wish to provide a few mathematical intuitions that might facilitate a deeper understanding of the presented concepts. By Eq.~\ref{dynamics}, the structure of the network determines its dynamic evolution over time. Mathematically, the structural connectivity matrix $\bm{A}$ serves as linear operator that maps each state, $\bm{x}$, to the rate of change from that state, $\dot{\bm{x}}$. This linear transformation can be described in terms of the evolutionary modes of the system consisting of the ${N}$ eigenvectors of $\bm{A}$ and their associated eigenvalues (Fig. \ref{fig:fig1}D). Each eigenvector of $\bm{A}$ can be imagined as an axis of the linear transformation which remains invariant over time. Thus, the eigenvectors reflect directions in the state space along which the system independently moves, each characterized by a specific pattern of brain region activity. Each eigenvalue, in turn, determines the rate of growth or decay along its associated eigenvector; that is, each eigenvalue determines how slow or fast the system grows or decays in the direction defined by the eigenvector. Thus, the eigenvalues control the temporal persistence of the set of supported modes of activity.
\\
\\
Especially for the interpretation of results, it is important to keep in mind that the dynamic model is relatively simple and relies on the assumptions of linearity, time invariance, and freedom from noise. Linearity implies that the system evolves linearly over time which is not an accurate reflection of extended dynamics in most neural processes. However, it has been shown that non-linear dynamics can be locally approximated by linear dynamics \cite{galan2008network, honey2009predicting}. Time invariance implies that the system’s response does not depend on the time point because both the structural network $\bm{A}$ and the control set $\bm{B}_\kappa$ are constant over time. This assumption likely holds true for short time scales but could be challenged by long-term structural reorganization, which has been observed across development and adulthood \cite{may2011experience,tang2017developmental,baum2017modular}. Freedom from noise implies that all properties of signal propagation are accounted for deterministically by the model. Yet, noise is a feature of neural signals at both small \cite{brinkman2016how,mlyarski2018adaptive,gai2010slope} and large time scales \cite{garrett2014brain,breakspear2011networks}. Nevertheless, it is customary and reasonable when first developing a mathematical model of a complex system to consider the salient features of the model that do not depend on noise \cite{kim2019linear,gollo2014mechanisms,taylor2014computational}.
\\

\subsection{Prerequisites}
The core of the dynamic model is the network structure that enables activity changes within a particular brain region to diffuse and induce state changes in connected regions. Thus, the first step is to build a structural connectivity network by defining the weighted adjacency matrix $\bm{A}$ (Fig. \ref{fig:fig1}A). The structural network of human and nonhuman animals can be modeled using a range of spatial scales of neural units and physical links between them \cite{bassett2018nature}. Here, we focus on the construction of the human connectome which requires (i) a brain parcellation that defines the ${N}$ nodes of the network and (ii) diffusion imaging data that define the strength of structural connectivity $A_{i,j}$ between two brain regions $i$ and $j$. To avoid self-loops in the system, we set the diagonal of $\bm{A}$ to zero. As a further practical note, the sparse nature of human connectomes typically does not require any thresholding of the matrix.
\\
\\
The next step is to pick a time-system that best reflects the neural dynamics under study. Here, we consider two options: discrete and continuous. A discrete-time system assumes that the system evolves in discrete time steps whereas a continuous-time system models continuously changing dynamics. Neural processes can often not be clearly assigned to one of these categories. The spatial scale of the neural unit under study is one factor that could guide this choice \cite{kim2019linear}. Smaller spatial scales such as single neurons could be modeled as either firing or not, and the accompanying uniform delays in neural activity could be better approximated by discrete-time systems. Larger neural units such as brain regions usually contain more neurons. The ensuing more heterogeneous timing of neural state changes might be better represented by continuous-time systems. Depending on the choice, the modeled dynamics can differ substantially because of their distinct mathematical implementation. More concretely, discrete-time dynamics rely on difference equations whereas continuous-time systems are based on differential equations. Note that we exclusively present formulas for continuous-time systems in the main text; the discrete-time versions can be found in the Supplementary Formulas.
\\
\\
The third step is to choose a method to stabilize the system to avoid its infinite growth over time. Because extremely large brain states are neurobiologically implausible, we normalize the system such that it either approaches the largest supported mode of activity or goes to zero over time:
\begin{equation}\label{key}
\bm{A_{norm} }=\frac{\bm{A}}{|\lambda(\bm{A})_{max}|+c}-\bm{I}
\end{equation}
Here, $\bm{I}$ denotes the identity matrix of size $N\times N$, and $|\lambda(\bm{A})_{max}|$ denotes the largest eigenvalue of the system. To normalize the system, we must specify the parameter $c$, which determines the rate of stabilization of the system. If $c=0$, the largest mode of activity is stable and all other modes decay; thus, the system approaches the largest mode over time. If $c>0$, such as the commonly used choice $c=1$, all modes decay; thus, the system goes to zero over time. As will become clear in the next section, the latter variant can be especially useful for the computation of average controllability in infinite time as well as modal controllability due to its mathematical definition. 
\\

\subsection{Optimal control energy} 
To quantify the degree of controllability of a network, we consider an optimal control problem to steer the network from a specific initial state $\bm{x}(0) = \bm{x}_0$ to a specific target state $\bm{x}(T) = \bm{x}_T$ over the time horizon $T$ while minimizing a combination of both the length of the state trajectory and the required control energy \cite{gu2017optimal, boltyanskii1960theory, hespanha2018linear}. Formally, we consider the problem
\begin{equation}\label{eq: optimal problem}
\bm{u}(t)_{\kappa}^{{\ast}} = \operatorname*{argmin}_{\bm{u}_{\kappa}} J (\bm{u}_{\kappa}) = 
\operatorname*{argmin}_{\bm{u}_{\kappa}}
\int_{0}^{T}((\bm{x}_{T}-\bm{x}(t))^{\top}(\bm{x}_{T}-\bm{x}(t)) + \rho \bm{u}_{\kappa}(t)^{\top}\bm{u}_{\kappa}(t))dt ,
\end{equation}
where the parameter $\rho$ determines the relative weighting between the costs associated with the length of the state trajectory and input energy. We use the cost function $J (\bm{u}(t)_{\kappa}^{{\ast}})$ to find the unique optimal control input $\bm{u}(t)_{\kappa}^{{\ast}}$ which allows us to calculate the \emph{optimal control energy} (Fig. \ref{fig:fig1}E) required by a single brain region $i$ (Fig. \ref{fig:fig1}C): 
\begin{equation}\label{key}
E_{i}^* = \int_{0}^{T} \lVert{u_{i}^{\ast}(t)\rVert^{2}_{2}}dt,
\end{equation}
and in total
\begin{equation}\label{key}
E^* = \sum_{i=1}^{N} E_{i}^* ~=\int_0^T\bm{u}_{\kappa}^*(t)^{\top}\bm{u}_{\kappa}^*(t)dt.
\end{equation}
To calculate optimal control energy, we must specify an initial brain state $\bm{x}_{0}$ and a target brain state $\bm{x}_{T}$ by assigning each brain region an initial and target activity level. If available, we can extract regional activity values directly from functional neuroimaging data such as electrocorticography or magnetic resonance imaging \cite{stiso2018white, cornblath2018context,zoller2019structural}, or we can use model-based estimates of task-related activation such as $\beta$ values from a general linear model \cite{braun2019brain}. Otherwise, we can also model brain states by artificially defining a subset of nodes to be active, such as brain regions belonging to the same cognitive system \cite{gu2017optimal, betzel2016optimally,cui2018optimization}. Additionally, we must specify the control set $\bm{B}_{\kappa}$, a set of brain regions into which we wish to inject signals. Theoretically, this choice can vary from controlling a single region to controlling the full brain. The choice of small- to medium-sized control sets, however, can lead to large numerical instabilities that accumulate and bias the results. As a rule of thumb, it is advisable to ensure that the numerical error does not exceed $10^{-6}$. To reduce the numerical error of the calculation, we can also define a relaxed control set $\bm{B}_{\kappa}$ by allowing large control input to control regions and small, random inputs to all other brain regions \cite{stiso2018white}. We must also specify the time horizon $T$ over which the control input is effective. For pragmatic reasons such as the potential translation to real external brain stimulation, the time horizon is usually set to finite time. Note that the time horizon is measured in arbitrary units even if brain states are defined by functional imaging data. Finally, we must specify the time step $dt$, which should be small enough to sufficiently approximate the dynamics \cite{gu2017optimal}; a reasonable choice is $dt=0.001$.
\\
\\
The cost function $J$ is motivated by the fact that biological systems might constrain the features of the traversed states, such as their type, diversity, or magnitude. Transitioning through states not too far away from the target state is supposed to avoid extremely large and thus neurobiologically implausible brain state transitions. In the case where no specific assumptions are made on the relative importance of the two constraints and where both the distance and energy values are of a comparable scale, an equal weighting of $\rho=1$ is a reasonable choice. Depending on our neurobiological assumptions, we can also define alternative cost functions and potentially restrict them to a subset of brain regions
\cite{cui2018optimization}.
\\

\subsection{Minimum control energy} 
A specific and commonly used subform of optimal control energy is obtained by letting $\rho \rightarrow \infty$ in \eqref{eq: optimal problem}, so that the cost function $J$ accounts only for the energy of the control input to steer the network from an initial state $\bm{x}(0) = \bm{x}_0$ to a target state $\bm{x}(T) = \bm{x}_T$. Thus, we call this metric \emph{minimum control energy} (Fig. \ref{fig:fig1}E). To compute the minimum control energy for a given network, it is convenient to define the controllability Gramian as
\begin{equation}\label{eq: gramian}
\bm{W}_{\kappa,T}=\int_{0}^{T}e^{\bm{A}t}\bm{B}_{\kappa}\bm{B}_{\kappa}^{\top}e^{\bm{A}^{\top}t}dt.
\end{equation}
The eigenvalues of $\bm{W}_{\kappa,T}$ can be used to answer several questions regarding the controllability of a network. First, if the smallest eigenvalue of $\bm{W}_{\kappa,T}$ is zero, then the network is not controllable. That is, there exist final states $x_T$ that cannot be reached by any control input, independent of its energy. Second, the magnitude of the smallest eigenvalue of $\bm{W}_{\kappa,T}$ is inversely proportional to the largest energy needed to reach a final state. That is, there exists a final state $\bm{x}_T$ that can be reached only using inputs whose energy is at least proportional to the inverse of the smallest eigenvalue of $\bm{W}_{\kappa,T}$. The foundational papers \cite{gu2015controllability, menara2018structural} have shown that brain networks are controllable from any single region; that is, the smallest eigenvalue of $\bm{W}_{\kappa,T}$ is greater than zero. However, brain networks require very large control energy; that is, the smallest eigenvalue of $\bm{W}_{\kappa,T}$ can be extremely small. It should also be noted that the computation of the smallest eigenvalue of $\bm{W}_{\kappa,T}$ tends to be numerically difficult, which motivates the next metric.
\\

\subsection{Average controllability} 
Apart from examining specific state transitions, the theoretical framework also allows us to ask questions regarding the general role of brain regions in controlling neural dynamics. A third metric is obtained by measuring the average input energy required to drive the system via a specified set of control nodes to all possible target states $\bm{x}_T$ with unit norm \cite{marx2004optimal, shaker2013optimal}. Following the above discussion, this metric equals $Trace (\bm{W}_{\kappa,T}^{-1})$ where the inverse of the controllability Gramian is a map from target states to control energy. To avoid numerical difficulties when controlling only a few nodes in very large systems, the metric is often approximated as $Trace (\bm{W}_{\kappa,T})$. Using this last form, our definition of \emph{average controllability} measures the ability of a network to amplify and spread control inputs, rather than being associated with the problem of steering the network state from $\bm{x}_0$ to $\bm{x}_T$. More concretely, average controllability quantifies the energy of the impulse response of a system, which describes how a system naturally evolves over time from some initial condition \cite{kailath1980linear}. Starting from an exclusive activation of the specified control regions, we observe the brain's natural response (Fig. \ref{fig:fig1}B). The larger and more variable this natural response, the more states can be reached with low energy input by controlling this specific set of brain regions. In prior work, average controllability was also intuitively described as the ability of a set of control nodes to drive the system to easily reachable, nearby states (Fig. \ref{fig:fig1}E) \cite{gu2015controllability}. 
\\
\\
To calculate average controllability, we must specify the time horizon $T$, which is the time period over which we wish to observe the impulse response of the system. Note that the units of the time horizon depend on the units of $\bm{A}$. To observe the complete impulse response, we often assume infinite time. Furthermore, we must determine the control set $\bm{B}_{\kappa}$, which is the set of brain regions into which control input can be injected. Even if the control set can comprise multiple, and even all nodes, average controllability is often examined for individual brain regions to enable comparison to another single-node metric: most commonly, modal controllability.
\\

\subsection{Modal controllability} 
Lastly, we introduce the metric \emph{modal controllability}, which was previously described as the ability of a single node to drive the system to distant, more difficult-to-reach states (Fig. \ref{fig:fig1}E) \cite{gu2015controllability}. The controllability metric is obtained directly from the eigenvalues and eigenvectors of the network weighted adjacency matrix. In particular, we use
\begin{equation}\label{key}
\phi_{i} = \sum^N_{j=1}(1 - (e^{\lambda_{j}(\bm{A})}))v_{ij}^{2},
\end{equation}
as a scaled summary of node $i$'s ability to control all ${N}$ modes of the network \cite{pasqualetti2014controllability}. To calculate modal controllability, we are not required to specify any parameters except the symmetric adjacency matrix $\bm{A}$. This metric capitalizes on information housed in the modes of $\bm{A}$, as summarized in the eigenvalues $\lambda_{j}$ and the matrix of normalized eigenvectors $\bm{V}=[v_{i,j}]$. Entry $v_{i,j}$ is a measure of the controllability of mode $\lambda_{j}(\bm{A})$ from node $i$ that geometrically corresponds to projecting node $i$ onto the eigenvector $j$ (Fig. \ref{fig:fig1}D) \cite{hamdan1989measures, kailath1980linear}. According to this heuristic, the larger the magnitude of the projection, the higher the ability of node $i$ to control mode $j$. The metric summarizes this notion across all modes, and then scales them by their rate of decline as determined by the eigenvalues. This weighting emphasizes especially fast decaying modes which might on average be more difficult to control because the injected control energy only has a short-term impact. We note that modal controllability is exclusively formalized for symmetric matrices whereas all of the other definitions that we present can be naturally extended to directed networks.
\\
\\
For completeness, we note that boundary controllability is another controllability metric used in the literature and measures the ability of a brain region to integrate information between network communities \cite{gu2015controllability, pasqualetti2014controllability}. However, because the metric is less commonly used, we will not discuss it further in this work.

\section{Materials and methods}

\subsection{Acquisition of diffusion imaging data}
High resolution anatomical brain images were collected from 10 healthy young adults (23.9 $\pm$ 3.6 years; 20-31 years; 70\% female). The participants underwent a 53:24 minute diffusion spectrum imaging (DSI) scan with 730 diffusion directions (maximum $b$-value = 5010s/mm$\textsuperscript{2}$, 21 b = 0 images, TR = 4300ms, TE = 102ms, matrix size = 144$\times$144, field of view = 260 $\times$ 260mm$\textsuperscript{2}$, slice number = 87, resolution = 1.8$\times$1.8$\times$1.8mm$\textsuperscript{3}$, multi-band acceleration factor = 3). Additionally, T1-weighted images were obtained using an MPRAGE sequence (TR = 2500ms, TE = 2.18ms, flip angle = 7 degrees, slice number = 208, slice thickness = 0.9mm). Both scans were acquired on a Siemens Magnetom Prisma 3 Tesla scanner with a 64-channel head coil. The study was approved by the Institutional Review Board of the University of Pennsylvania and all participants provided informed consent in writing.
\\

\subsection{Preprocessing of diffusion imaging data}
As previously described in more detail \cite{kim2018role}, the individual DSI scans were skull-stripped, realigned, and motion-corrected using an improved average b=0 reference image. The preprocessing was implemented in nipype \cite{Gorgolewski2011} using the Advanced Normalization Tools (ANTs, \cite{avants2011reproducible}) for image registration. We quantified the diffusion at different orientations in each voxel using the generalized $q$-sampling reconstruction method \cite{yeh2010generalized} in DSI Studio (dsi-studio.labsolver.org). Based on the derived quantitative anisotropy values, we performed deterministic tractography across the whole-brain \cite{yeh2013deterministic}. For each participant, we generated 1,000,000 streamlines with a maximum length of 500mm \cite{cieslak2014local} and a maximum turning angle of 35 degrees \cite{bassett2011conserved}.
\\

\subsection{Construction of structural brain networks}
Based on the diffusion imaging data, we constructed a structural brain network for each participant. Consistent with previous work \cite{gu2015controllability, gu2017optimal, betzel2016optimally, tang2017developmental}, we defined nodes of the network as brain regions according to the 234-node Lausanne atlas (excluding brainstem) \cite{cammoun2012mapping}. For this purpose, the Lausanne parcels were dilated by 4mm so that the parcels reached down into the white matter enough to ensure accurate sampling of underlying fibers. In the process of dilation, some voxels were assigned to two or more regions of interest; to eradicate this redundancy, we assigned each voxel to the mode of its neighbors \cite{daducci2012connectome}. After warping the parcellation into the subject's diffusion space, we quantified the edges of the network as total streamline count connecting a pair of brain regions, corrected for their volume. Overall, we constructed a 233$\times$233 sparse, weighted, and undirected adjacency matrix for each participant with the number of interregional streamlines representing structural connectivity.
\\

\subsection{Mapping to cognitive systems}
To define neurobiologically meaningful brain states, we capitalized on an established functional brain atlas \cite{yeo2011organization}. By clustering the resting state functional magnetic resonance imaging data of 1000 healthy adults, Yeo \emph{et al.} identified seven cognitive systems, each consisting of a set of distributed brain regions that are functionally coupled \cite{yeo2011organization}. The functional parcellation comprises visual (VIS), somatomotor (SOM), dorsal attention (DOR), ventral attention (VEN), limbic (LIM), frontoparietal control (FPC), and default mode (DM) systems. To link the functional and anatomical atlases, we mapped each brain region to the cognitive system with the highest spatial overlap as reported previously \cite{baum2017modular,cui2018optimization}. More concretely, each Lausanne parcel was assigned to the cognitive system that was most frequently associated with its voxels as defined by the purity index. Subcortical regions were summarized in an eighth, subcortical system (SC).
\\

\subsection{Probing different modeling choices}
We used the structural connectivity matrices of our sample to probe the impact of several modeling choices on average and modal controllability, and on minimum and optimal control energy. In our analyses, we systematically varied one parameter at a time while keeping all other parameters constant. Constant modeling choices were guided by the modeling choices most commonly used in the literature \cite{gu2015controllability, gu2017optimal, betzel2016optimally, tang2017developmental}. Concretely, we employed a simplified noise-free linear continuous-time and time-invariant network model, stabilized using $c=1$. When estimating average controllability, we set the time horizon $T$ to infinite time. When estimating control energies, we used $T=3$, approximated by 1000 time steps. When the system matrix $\bm{A}$ is stable, the controllability Gramian equation converges as $T$ approaches infinity. In this case, the Gramian can be computed algebraically by solving the Lyapunov equation.
\\

Motivated by the questions most relevant to each approach, we calculated average and modal controllability for each brain region based on single-node control sets, whereas control energies were based on full brain control. To examine control energies for transitions between previously defined functional systems, we simulated state transitions from an initially active default mode system to the activation of six different cognitive systems representing the target states \cite{yeo2011organization}. For each specific brain state, regions belonging to the activated cognitive system were set to one, whereas all other brain regions were set to zero. Except for the section on full versus partial control, we averaged across the examined state transitions. For optimal control energy, we set the relative energy weight $\rho=1$. In the restricted set of state transitions we investigated, minimum and optimal control energy yielded highly similar results. To avoid redundancy, we report the results on optimal control energy in the Supplementary Results (SFig. 1-6). Nevertheless, we point out deviating results of optimal control energy in the main text.
\\

\subsection{Examining metrics on an individual and regional level}
Since network control theory can be utilized to examine controllability differences in both individuals and brain regions, we separately studied the metrics on an individual and regional level. For this purpose, we summarized average controllability, modal controllability, and minimum control energy across either brain regions or individuals to subsequently investigate individual and regional values, respectively (Fig. \ref{fig:fig1}F). To estimate the consistency of a metric across parameter choices on an individual (regional) level, we first summarized the metric across brain regions (individuals) and then computed the Pearson correlation between the individual (regional) values obtained with one modeling choice and the individual (regional) values obtained with the second modeling choice; for example, we compare discrete- and continuous-time systems, and we compare two different parameter choices of time horizon $T$.
\\

\subsection{Construction of spatial adjacency network}
In addition to diffusing along white matter fibers, neural signals could potentially also diffuse between spatially adjacent brain regions. In other words, physical contact between two regions can be seen as a form of structural connectivity. To examine this complementary measure of structural connectivity, we generated brain networks, $\bm{S}$, based on the amount of shared neighborhood between two brain regions. We defined the edges of $\bm{S}$ as the number of face-touching voxels between two parcels of the Lausanne-atlas warped into subject space. In addition to studying each structural matrix separately, we also exploit the combined information of both measures by constructing the matrix $\bm{AS}$ as an average of $\bm{A}$ and $\bm{S}$. Because both diffusion and adjacency measures are expressed in arbitrary units and the actual scaling might impact controllability metrics, we scaled $\bm{S}$ and $\bm{AS}$ to the range of $\bm{A}$. We tested the effects of the structural connectivity types and their binarized version in a repeated measures ANOVA with two within-subject factors. To ensure that the effects of matrix type and binarization were not exclusively based on different edge weight distributions \cite{wu2018benchmarking}, we verified the results by sampling edge weights of $\bm{S}$ and $\bm{AS}$ from the distribution of $\bm{A}$ while preserving their rank order.
\\

\subsection{Controllability of fast and slow dynamics}
We capitalized on the concept of modal controllability to probe the ability of a brain region to control a specific set of temporal dynamics such as fast and slow modes \cite{stiso2018white, tang2019control}. Instead of summarizing across all modes that a system supports, we restricted the calculation of modal controllability to a subset of fastest (slowest) modes. We define transient (persistent) modal controllability as the ability of a brain region to control fast (slow) modes. The temporal dynamics of modes are determined by the magnitude of their eigenvalues. In continuous-time systems, large (small) eigenvalues relate to quickly (slowly) decaying modes. The lack of a formal definition of fast and slow dynamics requires the choice of a threshold that specifies the subset of modes (Fig. \ref{fig:fig1}D). We systematically probed the influence of threshold on a brain region's ability to control different temporal dynamics by calculating transient and persistent modal controllability using the 10\%, 20\%, 30\%, 40\%, and 50\% fastest and slowest modes. To disentangle these overlapping control tasks, we additionally summarized the ability of each brain region to control a specific interval of modes using the unscaled eigenvector matrix $\bm{V}$. For this purpose, we separated $\bm{V}$ into both 10 intervals and 2 intervals; this separation enabled a comparison to persistent and transient modal controllability based on a cut-off of 10\% and 50\%, respectively.
\\

\subsection{Definition of complexity of the energy landscape} 
The control trajectories from any initial state to any target state span the energy landscape of a dynamic system. The heterogeneity of the minimum control energy landscape is determined by the eigenvalues of the inverse of the controllability Gramian \cite{kailath1980linear}. Note that we assume independent control from all brain regions because the inverse of the Gramian is often ill-conditioned for small control sets \cite{gu2015controllability}. We capitalized on the variability of the eigenvalues to quantify the complexity of the minimum control energy landscape of a brain network, that is how the magnitude of the minimum control energy varies across all possible state transitions (Fig. \ref{fig:fig1}G). To account for the observed skewness of the distribution, we adopted the interquartile range as a measure of variability. Formally, we define the complexity of the energy landscape as the difference between the 75th and 25th percentile of the eigenvalue distribution of the inverted controllability Gramian
\begin{equation}\label{key}
C_{\kappa,T} = P_{75}(\lambda_{\bm{W}_{\kappa,T}^{-1}}) - P_{25}(\lambda_{\bm{W}_{\kappa,T}^{-1}}).
\end{equation}\\
We calculated the complexity of the energy landscape for each participant based on an infinite-time controllability Gramian. Then, we tested the complexity of the energy landscape of the brain network against three null models preserving distinct network characteristics. The topological null model preserved degree and strength distribution by iteratively switching connections between randomly selected edge pairs and subsequently associating the connections with the empirically observed edge weights \cite{rubinov2010complex}. The spatial null model preserved the relationship between Euclidean distance on the edge weights by adding the initially removed distance effects to the randomly rewired graph \cite{roberts2016contribution}. The combined null model preserved both the strength distribution and spatial embedding of the brain networks by approximating the observed strength distributions and effects of Euclidean distance on the edge weights \cite{roberts2016contribution}. Overall, we generated 1000 random instantiations of each null model.

\section{Results}

In the application of network control theory, we can rely on different neurobiological assumptions that are reflected in our modeling decisions. We begin with an examination of the impact of different modeling choices, before investigating several proposed model extensions.
\\

\subsection{Consistency across time systems}
\begin{figure*} [hbt!]
	\centering
	\includegraphics[width=0.95\textwidth]{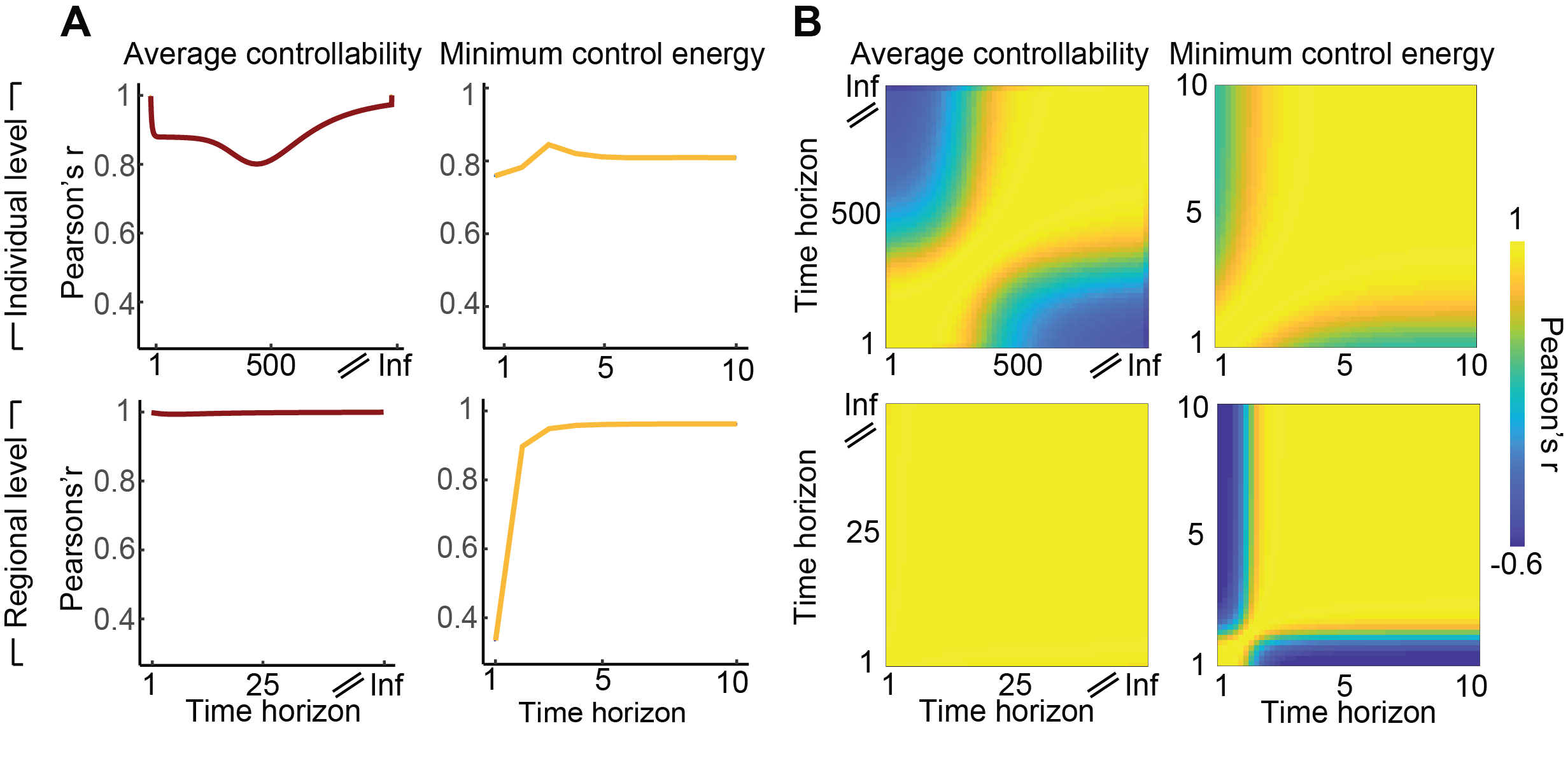}
	\caption{\textbf{Consistency of metrics across time.}
		\emph{(A)} Consistency of average controllability and minimum control energy across time systems. Pearson correlation coefficient between a given metric estimated for discrete- versus continuous-time systems across a range of time horizons $T$. \emph{(Left)} Average controllability; \emph{(Right)} minimum control energy. \emph{(B)} Consistency of average controllability and minimum control energy in a continuous-time system for any two choices of time horizon. Heat maps depict correlation matrix of different time horizons. Each heat map entry corresponds to the Pearson correlation of a metric based on two different time horizon choices. \emph{(Top)} Pearson correlation of individual metrics averaged across brain regions. \emph{(Bottom)} Pearson correlation of regional metrics averaged across participants. From these results, we deduce that average controllability and minimum control energy differ qualitatively for discrete- versus continuous-time systems when comparing estimates from short time horizons versus from longer time horizons.} 
	\label{fig:fig2}
\end{figure*}

When examining how the brain’s architecture gives rise to its complex dynamics by means of network control theory, one of the first modeling steps represents the type of the dynamic model. We can either assume that the neural dynamics evolve in discrete time steps or continuously. In light of potentially distinct dynamics of discrete- and continuous-time systems, we initially examined the consistency of minimum control energy, average controllability, and modal controllability across time systems. For this purpose, we calculated the Pearson correlation of each metric between discrete- and continuous-time systems, separately summarized across brain regions and individuals, and -- if applicable -- for different time horizons $T$ (Fig. \ref{fig:fig2}A). Average controllability showed a high consistency across time systems (individual level: $r_{min}=0.80$, $p=5\times 10^{-3}$; regional level $r_{min}=0.99$, $p=5\times 10^{-221}$), particularly for time horizons close to zero or infinity. Likewise, modal controllability demonstrated a high consistency across time systems (individual level: $r=0.99$, $p=3\times 10^{-11}$; regional level $r=1.0$, $p=2\times 10^{-16}$). Minimum control energy, however, was less consistent across time systems (individual level: $r_{min}=0.77$, $p=0.01$; regional level $r_{min}=0.33$, $p=2\times 10^{-7}$), particularly for short time horizons. The observed results are in line with theoretical considerations that suggest a convergence of discrete- and continuous-time systems for infinite time. Overall, the consistency across discrete- and continuous-time systems was high but depended on the metric, the observation level, and the chosen time horizon.
\\

\subsection{Consistency across time horizons}
Network control theory might lend itself particularly well to evaluate how local perturbations of the brain, for instance elicited by deep brain stimulation or transcranial magnetic stimulation, affect whole brain dynamics. In such a setting we might be interested in assessing different temporal scales of brain stimulation such as the effect of stimulation in the short term or in the long term. This question prompts the examination of the time horizon of the injected signal as another early modeling decision. We addressed this question by quantifying the Pearson correlation between values estimated for one time horizon $T$ and for another time horizon $T'$, separately averaged across brain regions or across individuals (Fig. \ref{fig:fig2}B). We first noted that the time horizon affected the scaling of the metrics (SFig. 1A). More specifically, average controllability monotonically increased in magnitude with larger time horizons because we observed the impulse response of the system for a longer time interval. Minimum control energy monotonically decreased with larger time horizons; this relation is intuitive when we consider the fact that longer time horizons allow the system to capitalize on its own natural dynamics, thereby demanding less exogenous control input. In contrast, optimal control energy first rapidly decreased and then slightly increased with larger time horizons (SFig. 1A). The increasing amount of optimal control energy might be required to additionally constrain the distance of traversed brain states over longer time horizons. In general, we found a high consistency between the metrics across a wide range of examined time horizons. However, smaller time horizons demonstrated a different control regime in which average controllability (individual level: $r_{min}=-0.56$, $p=0.09$; regional level ($r_{min}=0.88$, $p=7\times 10^{-78}$) and minimum control energy (individual level: $r_{min}=0.28$, $p=0.44$; regional level ($r_{min}=-0.71$, $p=6\times 10^{-37}$) were partly anti-correlated with the corresponding metrics in larger time horizons. In sum, short time horizons induced an alternative control regime in average controllability and minimum control energy compared to longer time horizons.
\\

\subsection{Impact of normalization}
\begin{figure*} [hbt!]
	\centering
	\includegraphics[width=0.95\textwidth]{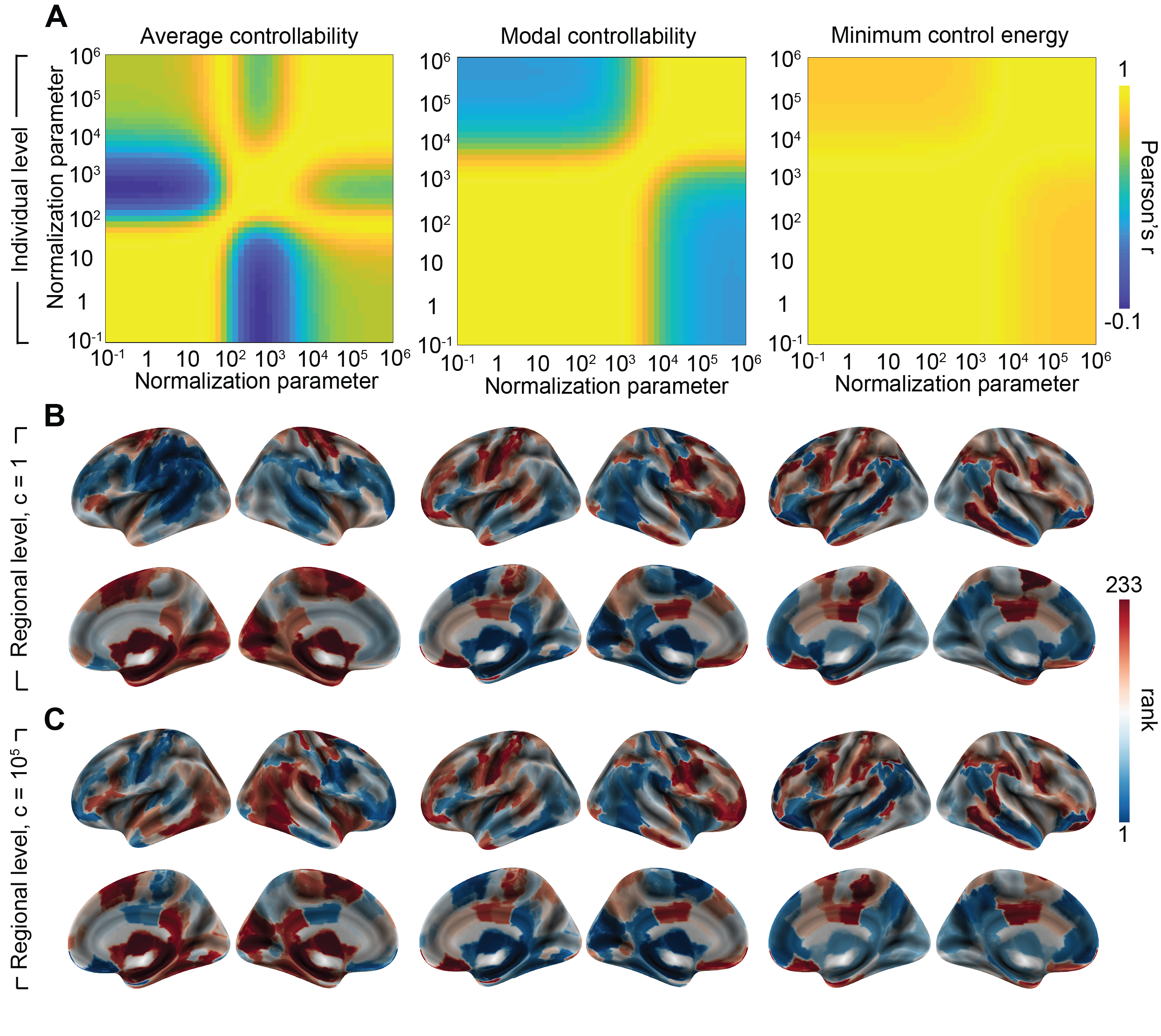}
	\caption{\textbf{Different control regimes depending on normalization.} Consistency of \emph{(Left)} average controllability, \emph{(Middle)} modal controllability, and \emph{(Right)} minimum control energy for different choices of the normalization parameter $c$. \emph{(A)} Heat maps depict correlation matrices of different normalization parameters. Each heat map entry corresponds to the Pearson correlation between a metric calculated using one normalization parameter and the same metric calculated using a second normalization parameter. Pearson correlation of individual metrics summarized across brain regions. Normalizing the system such that it stabilizes faster introduces a different control regime. \emph{(B)} Regional controllability metrics and control energy projected onto the brain surface using a normalization parameter $c=1$. \emph{(C)} Analogously, the same metrics plotted onto the brain surface using a normalization parameter $c=10^{5}$. Together, panels \emph{(B)} and \emph{(C)} illustrate the fact that distinct choices for the normalization parameter can induce distinct control regimes. Note: metric values are ranked for visualization purposes only.}
	\label{fig:fig3}
\end{figure*}

The normalization step represents another modeling decision that is related to time. For mathematical reasons, we often assume the neural dynamics to diminish and stabilize over time. Neurobiological considerations determine the degree of normalization; that is, how fast or slow we assume the neural system to stabilize. To investigate the effect of normalization on controllability metrics and control energies, we calculated average controllability, modal controllability, and minimum control energy for different choices of the normalization parameter $c$. At both individual and regional levels, we first observed that with increasing $c$, average controllability decreased whereas modal controllability and minimum control energy increased (SFig. 2). Next, we investigated the consistency of the metrics across different manners of normalization by quantifying the Pearson correlation between metrics for two choices of the normalization parameter $c$, separately summarized across brain regions (Fig. \ref{fig:fig3}A) and individuals (SFig. 3A). In both cases, we observed two different control regimes depending on small ($c=0.1$ to $c=10^{2}$; Fig. \ref{fig:fig3}A) and large ($c=10^{4}$ to $c=10^{6}$; Fig. \ref{fig:fig3}C) normalization parameters. Within each regime, the results were highly consistent independent of the normalization parameter $c$. Between both regimes, however, the consistency in average controllability (individual level: $r_{min}=-0.19$, $p=0.61$; regional level: $r_{min}=0.86$, $p=1 \times 10^{-69}$;), modal controllability (individual level: $r_{min}=0.29$, $p=0.41$; regional level: $r_{min}=0.99$, $p=7 \times 10^{-320}$), and minimum control energy (individual level: $r_{min}=0.87$, $p=2 \times 10^{-3}$; regional level: $r_{min}=0.81$, $p=6 \times 10^{-56}$) was reduced. Hypothesizing that this alternative control regime might be due to a faster stabilization of the system, we quantified the Spearman correlation between $c$ and the decay rate of the slowest mode. We indeed found that an increase of the normalization parameter led to a faster stabilization of the system ($r=-1.0$, $p=0$). Taken together, a faster stabilization of the system introduced an alternative control regime that particularly affected controllability metrics.
\\

\subsection{Impact of control set size}
\begin{figure*} [hbt!]
	\centering
	\includegraphics[width=0.95\textwidth]{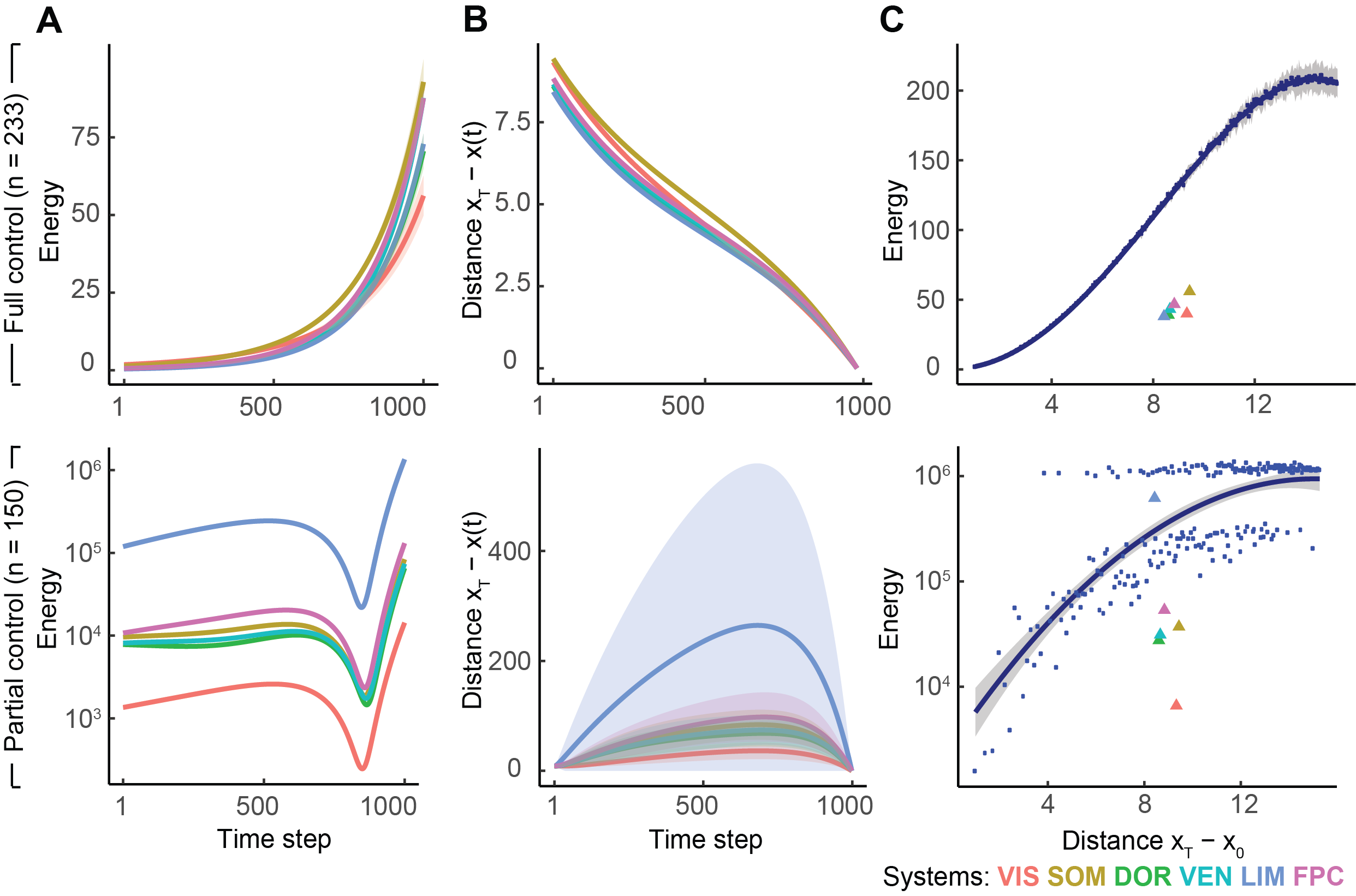}
	\caption{\textbf{Minimum control energy in full and partial control sets.} Minimum control energy for state transitions from the default mode system to six target cognitive systems (colored lines and triangles). \emph{(Top)} Results from simulations using a whole-brain control set including all 233 brain regions. \emph{(Bottom)} Results from simulations using a control set consisting of random subsets of 150 brain regions. \emph{(A)} Minimum control energy across the control trajectory differs quantitatively between full and partial brain control. \emph{(B)} Euclidean distance between the current state and the target state across the control trajectory differs between full and partial brain control. \emph{(C)} Minimum control energy increases with larger Euclidean distance between initial and target states. Blue dots depict state transitions from a zero-activity brain state to states comprised of a varying number of randomly activated brain regions. All values are averaged across participants. Lines and ribbons represent the best fit to the data and the 95\% confidence interval, respectively. Abbreviations: $\bm{x}_{0}$ = initial state, $\bm{x}_{T}$ = target state, $\bm{x}(t)$ = state at time $t$, VIS = visual, SOM = somatomotor, DOR = dorsal attention, VEN = ventral attention, LIM = limbic, and FPC = frontoparietal control.}
	\label{fig:fig4}
\end{figure*}

In the study of the effects of brain stimulation on brain activity, we can also ask how many and which brain regions we should control in order to drive the system to a, for instance healthy, state. More concretely, we could compare the effects of targeting a specific neural circuit to the effects of whole-brain stimulation. This motivates the examination of a final modeling choice: the number of controlled brain regions. To probe the effect of control set size on minimum control energy, we began by generating random control sets of varying number of active brain regions ranging from single-node to full-brain control. We then proceeded by testing the impact on minimum control energy and the numerical error in six brain state transitions: from activation of the default mode to activation of six canonical cognitive systems as defined by Yeo \emph{et al.} \cite{yeo2011organization}. Importantly, the numerical error was reasonably small ($<1 \times 10^{-6}$) when we controlled at least 28.3\% to 29.6\% of brain regions ($N_{VIS}=66$, $N_{SOM}=67$, $N_{DOR}=67$, $N_{VEN}=68$, $N_{LIM}=69$, $N_{FPC}=68$), increasing our confidence in the results. We observed that minimum control energy and the numerical error decreased exponentially with increasing control set size (SFig. 4A). Intuitively, the control of a larger number of brain regions required less control energy. The exponential relationship between control energy and control node set can also be mathematically derived \cite{pasqualetti2014controllability}.
\\
\\ 
Next, we were interested in how control and state trajectories differ in partial- compared to full-brain control sets. We calculated the minimum control energy trajectory and the distance between the state trajectory and the target state for the same six state transitions controlling all versus randomly drawn sets of 150 brain regions. In full-brain control, we observed an exponential increase in energy (Fig. \ref{fig:fig4}A) and an approximately linear decrease in the distance between current and target state (Fig. \ref{fig:fig4}B) across the control horizon. When we controlled only a part of the brain, control and state trajectories differed considerably. For instance, instead of taking the direct route through the state space, the system traversed more distant states before it reached the target state. Theoretical work has indeed shown that such non-local trajectories generally emerge if only a subset of nodes is controlled \cite{sun2013controllability}.
\\
\\ 
Finally, we wished to study the effect of distance between initial and target state on minimum control energy. Because the set of state transitions that we studied lacked sufficient variability in these distances, we additionally simulated trajectories from a zero-activity initial state to random target states with a varying size of brain regions activated. We found a monotonic increase of minimum control energy with increasing distance between initial and target states (Fig. \ref{fig:fig4}C). When employing a partial control set, a subset of the random state transitions required massive amounts of control energy. A further exploration revealed that these hardly controllable state transitions involved an activation of two weakly connected limbic regions that were not part of the random control set. Similarly, the six state transitions likely required less control on average because the activation of densely connected cognitive systems is an easier control task than the activation of randomly chosen regions in target states of equal distance. The findings in optimal control energy were highly similar, even if the exact control and state trajectories were different (SFig. 5). Overall, state and control trajectories differed substantially depending on which brain regions were allowed to receive energy input.
\\

\subsection{Relation of metrics}
\begin{figure*} [hbt!]
	\centering
	\includegraphics[width=0.95\textwidth]{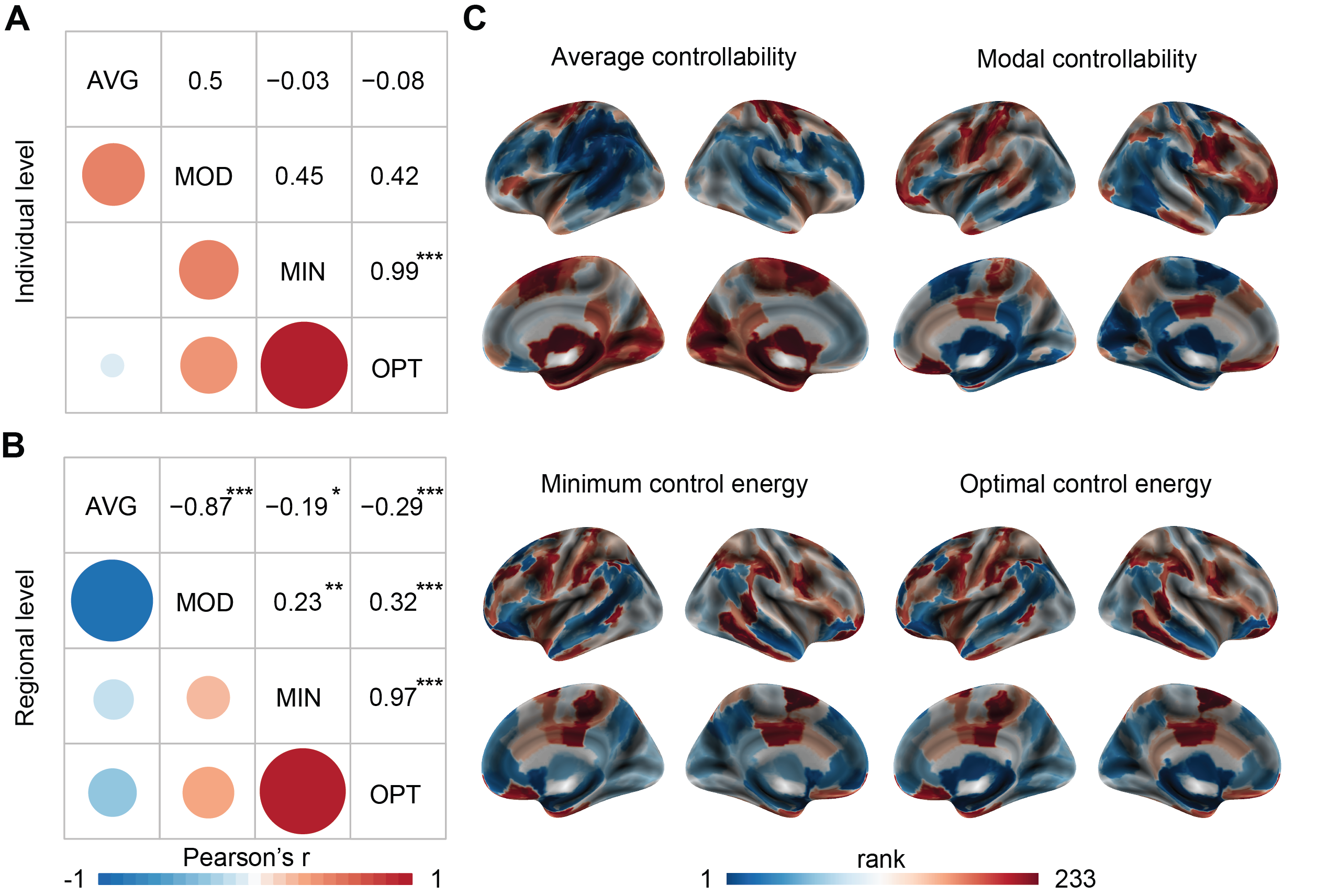}
	\caption{\textbf{Relation between controllability metrics and control energies.}
		\emph{(A)} The correlation matrix of individual average controllability, modal controllability, minimum control energy, and optimal control energy, summarized across brain regions. Pearson correlation coefficients (upper matrix triangle) additionally encoded via color and size of circles (lower matrix triangle). \emph{(B)} Analogously, the correlation matrix of regional controllability metrics and control energies averaged across participants. \emph{(C)} Regional controllability metrics and control energies projected onto the brain surface. Note that the metric values are ranked for visualization purposes only. Collectively, these panels illustrate the negative association between regional average and modal controllability, and the high consistency between minimum and optimal control energy. Abbreviations: AVG = average controllability, MOD = modal controllability, MIN = minimum control energy, and OPT = optimal control energy. Asterisks indicate significance on a Bonferroni-corrected $\alpha$-level of 0.05 (*), 0.01 (**), and 0.001 (***).}
	\label{fig:fig5}
\end{figure*}

How easily a brain network can be steered to different states and the amount of energy required to achieve a specific state transition could prove to be informative markers for pathology in or injury to the central nervous system. The selection of the metric put to the test primarily depends on the specific research question. To elucidate the empirical associations between controllability metrics and control energies, we measured the Pearson correlation between each pair of metrics, summarized separately across brain regions (Fig. \ref{fig:fig5}A) and individuals (Fig. \ref{fig:fig5}B and Fig. \ref{fig:fig5}C). In line with previous research \cite{gu2015controllability}, average and modal controllability showed a positive association on the individual level ($r=0.5$, $p=0.14$), but a significant large negative association on the regional level ($r=-0.87$, $p=1 \times 10^{-74}$). In the six state transitions we studied, we found a significant large positive correlation between minimum and optimal control energy both on the individual level ($r=0.99$, $p=1 \times 10^{-7}$) and on the regional level ($r=0.97$, $p=1 \times 10^{-136}$). Consistent with the mathematical constraints of their definitions, minimum control energy was significantly lower than optimal control energy on both the individual ($mean_{min}=43.8$, $mean_{opt}=56.0$, $V=0$, $p=2\times 10^{-3}$) and regional ($mean_{min}=0.19$, $mean_{opt}=0.24$, $V=2$, $p=6 \times 10^{-40}$) levels. Finally, we investigated how controllability metrics and control energies are related. On both levels of examination, average controllability demonstrated small negative correlations with control energies (ranging from $r=-0.29$, $p=6 \times 10^{-6}$ to $r=-0.03$, $p=0.93$), whereas modal controllability showed slightly larger, positive associations with control energies (ranging from $r=0.23$, $p=3\times 10^{-4}$ to $r=0.45$, $p=0.19$). The size of the correlations estimated with empirical data supported the scarcity of a clear theoretical link between the concepts. Yet, the directions of the effects was consistent with the general notion that high average controllability implies low control energy, whereas modal controllability is linked to higher control energies.
\\

\subsection{Structural connectivity measures}
\begin{figure*} [hbt!]
	\centering
	\includegraphics[width=0.95\textwidth]{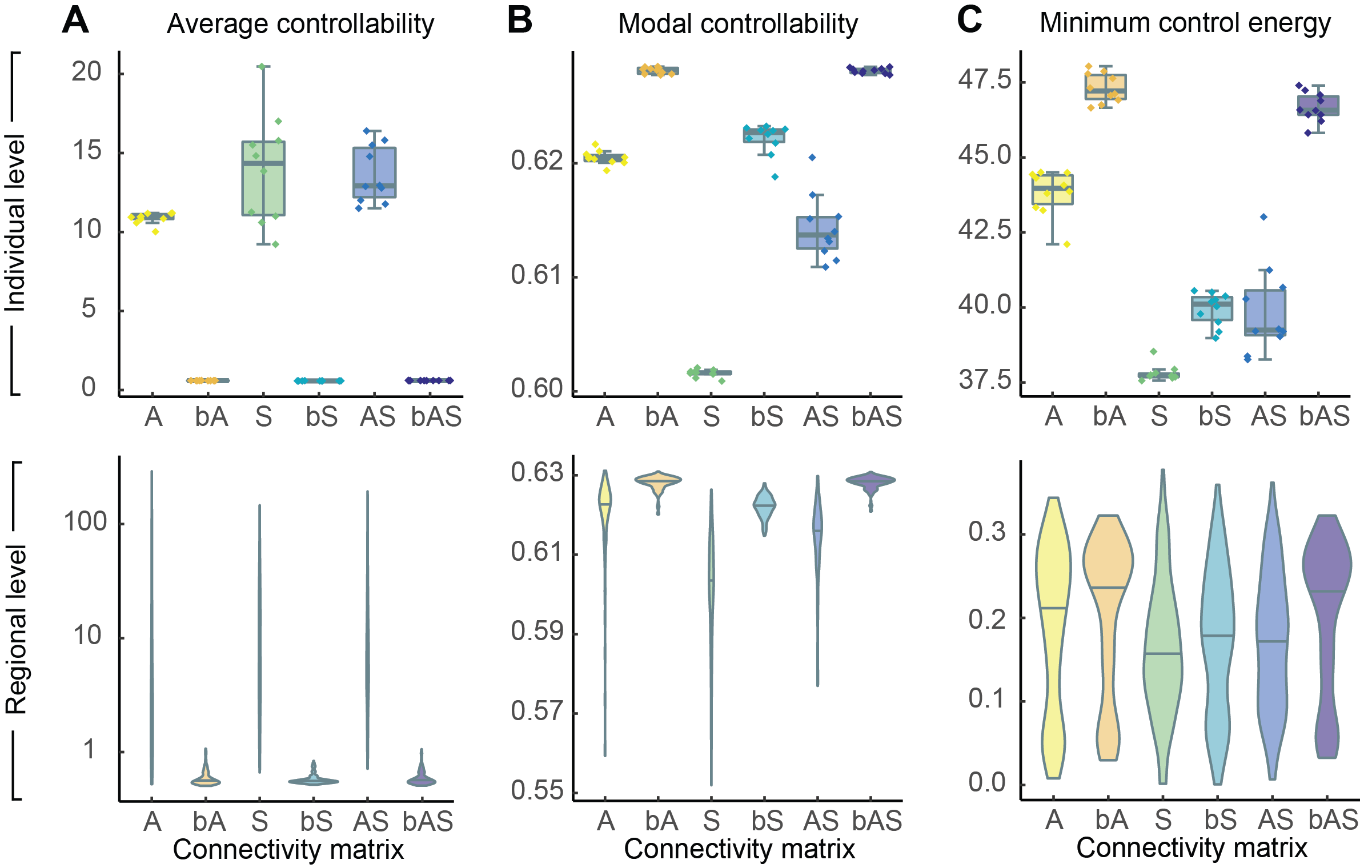}
	\caption{\textbf{Structural connectivity measures.}
		\emph{(A)} Average controllability, \emph{(B)} modal controllability, and \emph{(C)} minimum control energy for different measures of structural connectivity. The network encoded in $\bm{A}$ is based on streamline counts between two brain regions from diffusion imaging. The network encoded in $\bm{S}$ is based on the extent of spatial adjacency between two brain regions from T1-weighted images. The network encoded in $\bm{AS}$ is an average of $\bm{A}$ and $\bm{S}$. Additionally, we consider binary versions of the three networks, and refer to them as $\bm{bA}$, $\bm{bS}$, and $\bm{bAS}$, respectively. \emph{(Top)} Box plots depict individual controllability metrics and control energy summarized across brain regions. Diamonds represent individuals. \emph{(Bottom)} Violin plots depict regional metrics averaged across participants. Collectively, these panels illustrate the fact that the two structural connectivity measures provide complementary information that is retained by their combination.}
	\label{fig:fig6}
\end{figure*}

After systematically examining the impact of diverse modeling choices, we wished to provide several, potentially useful extensions of the theoretical framework. We begin with a consideration of the architecture of the brain which represents the core of network control theory. Thus, it is particularly relevant how we define the inter-connections between brain regions. Typically used DTI data do not take into account the fact that the signal can theoretically diffuse via physical contact between two brain regions. To evaluate the consequences of different forms of the adjacency matrix reflecting different modes of signal propagation in the brain, we additionally built structural connectivity networks based on the amount of shared neighborhood between two brain regions. Then, we calculated controllability metrics and control energies for the two alternative measures of structural connectivity, their combination, and their binarized versions (Fig. \ref{fig:fig6}). We first examined the similarity in controllability of structural networks based on diffusion imaging ($\bm{A}$) and based on spatial adjacency ($\bm{S}$). Between $\bm{A}$ and $\bm{S}$, we found small- to medium-sized Pearson correlations in average controllability (individual level: $r=0.02$, $p=0.95$; regional level: $r=-0.01$, $p=0.92$), modal controllability (individual level: $r=-0.15$, $p=0.67$; regional level: $r=0.41$, $p=10 \times 10^{-11}$), and minimum control energy (individual level: $r=0.36$, $p=0.31$; regional level: $r=0.64$, $p=2 \times 10^{-16}$). Thus, the two measures of structural connectivity provide complementary information. Next, we quantified the effect of binarization, matrix type ($\bm{A}$ vs. $\bm{S}$), and their combination $\bm{AS}$, on controllability metrics and control energy. Repeated measures ANOVAs revealed significant main effects of matrix type and binarization on a Bonferroni-corrected level of $\alpha=0.01$ except for the effect of matrix type on average controllability (individual level: $F=5.98, p=5 \times 10^{-3}$; regional level $F = 1.30$, $p=0.27$).
\\
\\
To ensure that these results were not exclusively due to different edge weight distributions, we verified these results using $\bm{S}$ and $\bm{AS}$ based on the same edge weight distribution as $\bm{A}$. When we examined the effects in more detail, we observed that the binarization reduced the absolute values and variance of average controllability on both the regional and individual level, whereas modal controllability displayed a reverse effect. This pattern of results is in line with findings that less connected brain regions exhibit lower average controllability but higher modal controllability \cite{betzel2016optimally}. Similarly, individual minimum control energy was increased for binary matrices compared to fully weighted matrices; this result is consistent with previous evidence demonstrating that control nodes with more homogeneous edge weights require larger control energy \cite{kim2018role}. Overall, the binarization of the structural connectivity matrix substantially reduced the variance of controllability metrics but not minimum control energy, suggesting that the edge weights carry valuable information especially for controllability metrics.
\\

\subsection{Persistent and transient modal controllability}
\begin{figure*} [hbt!]
	\centering
	\includegraphics[width=0.95\textwidth]{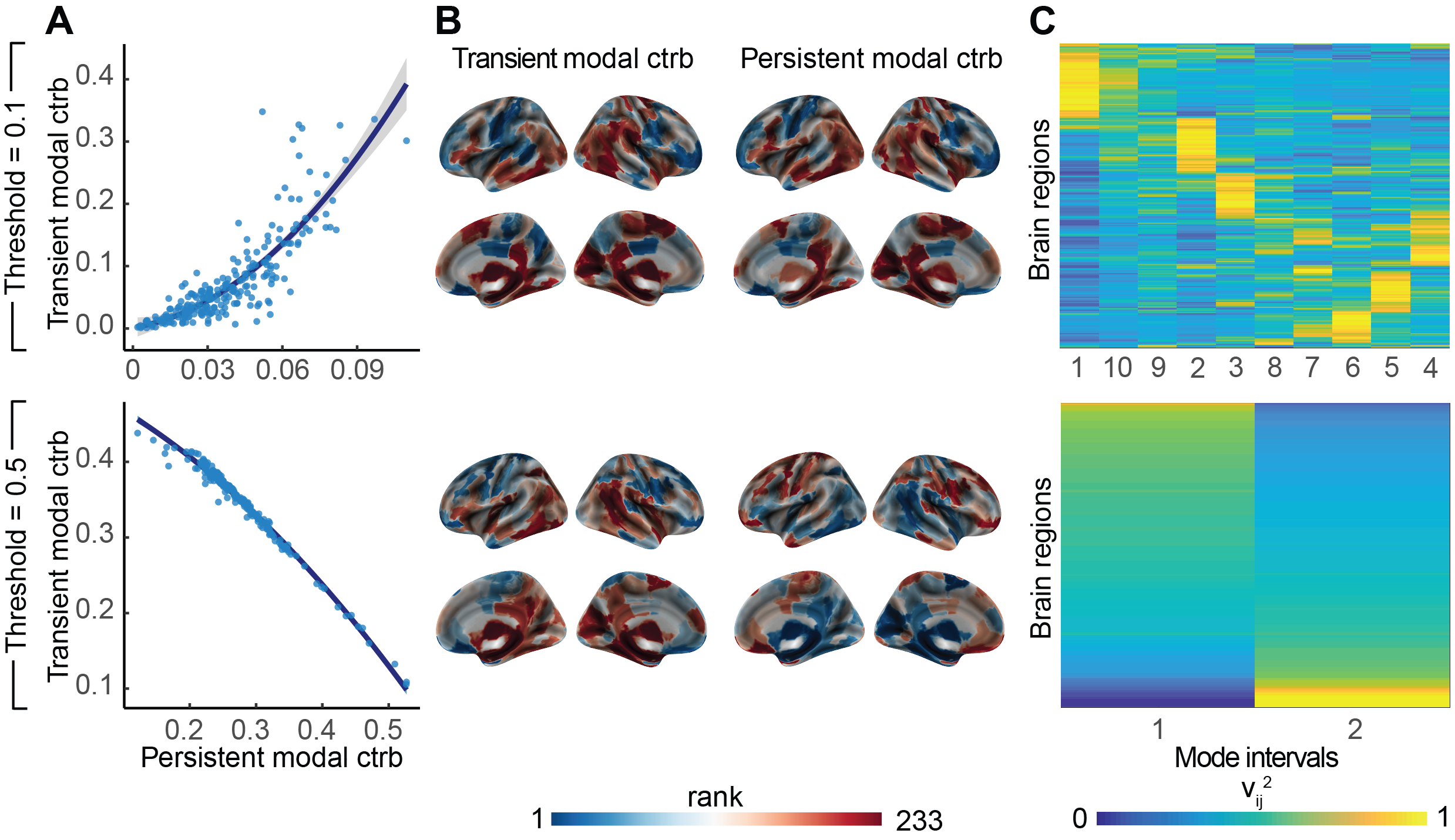}
	\caption{\textbf{Impact of threshold on persistent and transient modal controllability.} Regional controllability of fast and slow modes for two exemplary thresholds. \emph{(Top)} Persistent and transient modal controllability defined as a brain region's ability to control the 10\% slowest and 10\% fastest modes, respectively. \emph{(Bottom)} Analogously, persistent and transient modal controllability based on a threshold of 50\% of the modes. \emph{(A)} Scatter plots show the relationship between transient and persistent modal controllability of brain regions averaged across participants. \emph{(B)} Transient and persistent modal controllability projected onto the brain surface. Note that metric values are ranked for visualization purposes only. \emph{(C)} Heat maps depict each node's ability to control a specific interval of modes, ranging from the fastest (1) to the slowest (10 and 2 respectively) modes. For this purpose, we summarized the unscaled controllability matrix {\textit{V}} into 10 and 2 intervals respectively. When we aggregated the modes into 10 intervals, similar brain regions were capable of controlling both the slowest and fastest group of modes. When we, however, aggregated the modes into 2 intervals, brain regions were able to control either fast or slow modes. Thus, the ability of a brain region to control fast and slow modes depended on the definition of the specific control task. Abbreviations: ctrb = controllability, VIS = visual, SOM = somatomotor, DOR = dorsal attention, VEN = ventral attention, LIM = limbic, FPC = frontoparietal control, DM = default mode network, and SC = subcortical.}
	\label{fig:fig7}
\end{figure*}

Many neuroscientific endeavors focus on the speed of neural dynamics. Network control theory allows us to explicitly study whether a brain region is capable of controlling fast and slowly changing activity modes by means of transient and persistent modal controllability. However, there is no clear definition of which activity modes are considered as fast or slow. Thus, we wished to further inspect how the definition of fast and slow temporal dynamics affects transient and persistent modal controllability. We began with the calculation of both metrics across various thresholds for determining which modes were considered to be transient versus persistent. First, we observed that with increasing threshold the magnitude of both transient and persistent modal controllability increased because the number of summed modes was expanded. As expected, we further noted that transient and persistent modal controllability based on a threshold of 0.5 summed up to modal controllability. The initially positive Pearson correlation between transient and persistent modal controllability of brain regions reduced and turned into a negative association with increasing thresholds ($r_{0.1}=0.82$, $p_{0.1}=7 \times 10^{-59}$; $r_{0.2}=0.78$, $p_{0.2}=7 \times 10^{-49}$; $r_{0.3}=0.65$, $p_{0.3}=1 \times 10^{-29}$; $r_{0.4}=-0.20$, $p_{0.4}=3 \times 10^{-3}$; $r_{0.5}=-0.99$, $p_{0.5}=4 \times 10^{-187}$) (Fig. \ref{fig:fig7}A). Notably, for small thresholds such as 0.1, a subset of brain regions was found to be capable of controlling both fast and slow temporal dynamics (Fig. \ref{fig:fig7}B). While controlling for the size of each cognitive system, we found that these brain regions belonged primarily to the subcortex (36\%) and VIS (22\%) systems, but also VEN (12\%), DOR (9\%), SOM (8\%), DM (8\%), and FPC (5\%) systems. For large thresholds such as 0.5, brain regions seem to be either able to control fast dynamics (39\% SC, 14\% DOR, 12\% VIS, 11\% DM, 8\% FPC, 6\% SOM, 5\% VEN, and 5\% LIM systems) or slow dynamics (31\% FPC, 26\% subcortex, 22\% SOM, 10\% DOR, 7\% DM, and 3\% VIS systems), but not both.
\\
\\
To explore this ambiguous relationship in more detail, we disentangled the overlapping thresholds by considering the unscaled controllability matrix $\bm{V}$, and then by summarizing the modes into 10 intervals versus 2 intervals (Fig. \ref{fig:fig7}C, top versus bottom). Interestingly, this investigation into the controllability of separate mode intervals also supported the notion that similar brain regions were capable of controlling fast and slow dynamics in the strict definition of these control tasks (10 intervals) but not in the broader definition of these control tasks (2 intervals). Importantly, we note that these results do not extend to discrete-time systems because the definition of modes that are considered as fast versus slow differs substantially between time systems. Overall, the ability of a brain region to control fast and slow modes largely depended on the definition of the control tasks.
\\

\subsection{Complexity of energy landscape}
\begin{figure*} [hbt!]
	\centering
	\includegraphics[width=0.5\textwidth]{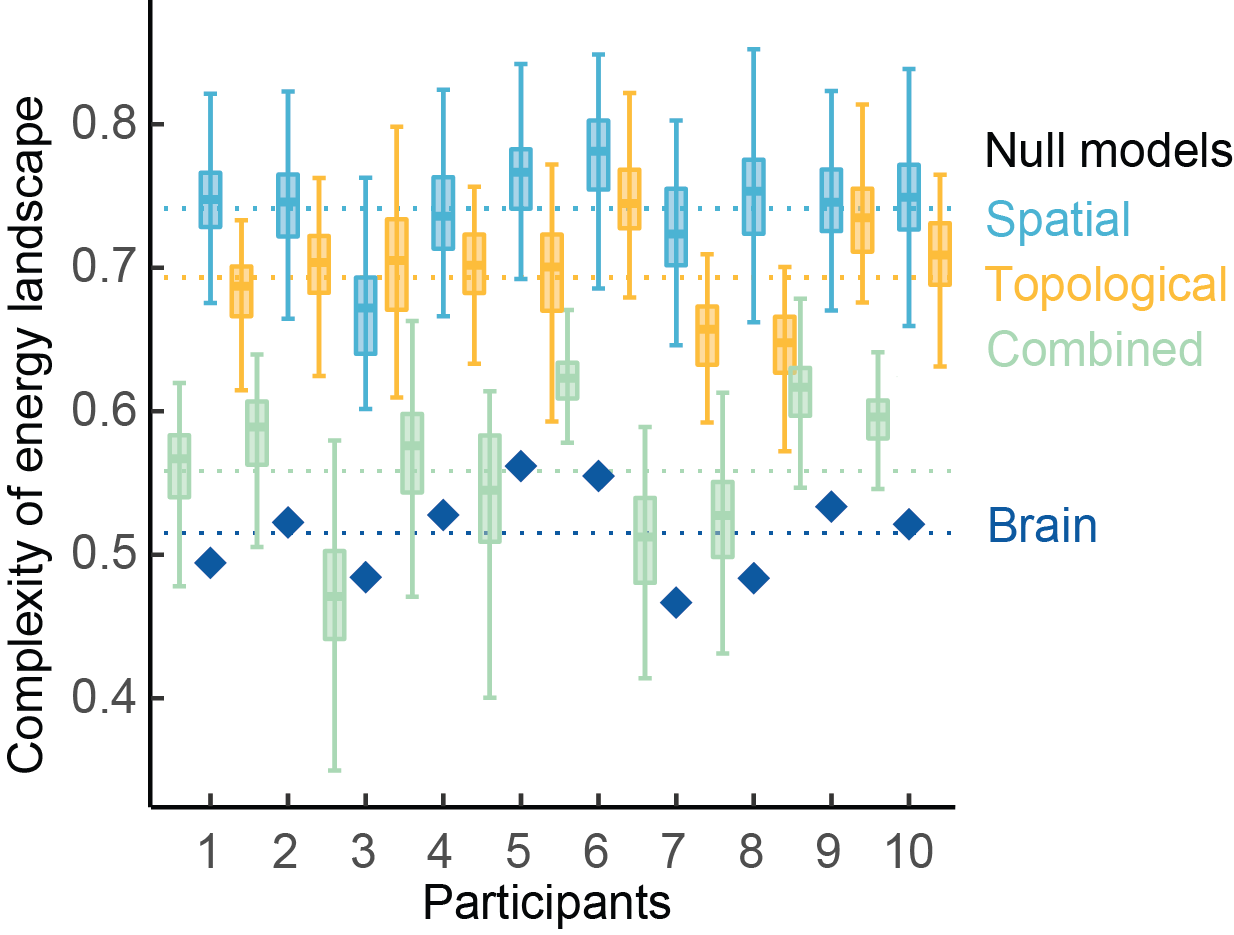}
	\caption{\textbf{Complexity of the energy landscape of the human brain.}
		Heterogeneity of the minimum control energy landscape of individual participants (dark blue diamonds) as compared to three null models preserving different characteristics of brain networks. The complexity of the energy landscape was quantified by the variability of the eigenvalue distribution of the controllability Gramian. Null model distributions (box plots) were estimated by randomly rewiring each brain network 100 times. Spatial null models (blue box plots) preserved the relationship between edge weight and Euclidean distance. Topological null models (yellow box plots) preserved degree and strength distributions. Combined null models (green box plots) preserved both strength distribution and spatial embedding. Dashed lines indicate complexity of the energy landscape of brain networks and null models averaged across individuals. The combination of topological and spatial characteristics partially explains the homogeneous energy landscape of the brain.}
	\label{fig:fig8}
\end{figure*}

Finally, we sought to extend the types of research question we can address with the set of currently available controllability and energy metrics. For this purpose, we developed and validated a complementary metric that measures the heterogeneity of all possible minimum control energy trajectories. The complexity of the energy landscape allows us to quantify the similarity or dissimilarity of all possible state transitions in respect to their required amount of control energy. Based on the variability of the eigenvalues of the controllability Gramian, we quantified the complexity of the minimum control energy landscape in each individual. Probing the consistency of the complexity of the energy landscape across time systems, we observed a large positive Pearson correlation between discrete- and continuous-time systems ($r=0.87$, $p=1\times 10^{-3}$). We further examined the complementarity of the complexity of the energy landscape by calculating the Pearson correlation between the complexity measure and the other established control metrics defined earlier. We found a small negative association between complexity and average controllability ($r=-0.15$, $p=0.68$), a large negative association with modal controllability ($r=-0.67$, $p=0.04$), and a medium negative association with minimum control energy ($r=-0.40$, $p=0.26$). Next, we validated the complexity of the energy landscape of the brain against three null models, preserving either the strength distribution or the spatial embedding, or both. Brain networks showed a significantly lower complexity of the energy landscape than the topological null model ($W=65$, $p=8\times 10^{-8}$), the spatial null model ($W=0$, $p=5\times 10^{-8}$), and the combined null model ($W=2498$, $p=6\times 10^{-3}$), as quantified by a Wilcoxon test (Fig. \ref{fig:fig8}). Interestingly, the combination of topological and spatial characteristics seemed to partially explain the brain's higher homogeneity of the energy landscape. We found consistent evidence in discrete-time systems (SFig. 7). Overall, the complexity of the energy landscape of the brain was complementary to other controllability metrics and low compared to several null models.
\\

\section{Discussion}

Network control theory is an emerging field in neuroscience that has the potential to yield promising insights into structure-function relationships in health and disease. Here, we provided an overview of the theoretical framework by illustrating the underlying model of neural dynamics and commonly studied controllability concepts. Based on the structural brain networks from ultra high-resolution diffusion imaging data (730 diffusion directions) of 10 healthy adults, we calculated average and modal controllability as well as minimum and optimal control energy. We then systematically probed the impact of different modeling choices, specifically the choice of time system, time horizon, normalization, and size of the control set, on these metrics. We further suggested potentially useful model extensions such as an alternative measure of structural connectivity accounting for propagation of signals through gray matter to abutting regions, and a complementary metric quantifying the complexity of the energy landscape of brain networks.
\\
\\
\textbf{Specific modeling recommendations.}
Based on our systematic examination of different modeling choices, we derived several specific recommendations. First, we observed a generally high consistency between the behavior of discrete- and continuous-time systems, which depended on the metric, observation level, and time horizon. Classifying the neural dynamics under study as clearly discrete- or continuous-time is often challenging. Unless an investigator has a clear justification for choosing one time system over another, we recommend to verify the obtained results in the alternative time-system to allow for a better generality of the findings and inferences drawn therefrom. Second, we demonstrated that short time horizons led to an alternative time system compared to longer time horizons. The arbitrary units of the time scale further challenge the decision of which time horizon to choose. If there exists no concrete justification for the choice of time horizon, we recommend to validate the obtained findings using several different time horizons. Third, we found that a fast stabilization of the system induced a substantially different control scenario. Again if there are no concrete neurobiological variables that can be used to constrain one's choice, we suggest that a slow stabilization could be a plausible representation of most neural dynamics, allowing for a broader range of dynamics. Since the influence of the normalization parameter $c$ depends on the largest eigenvalues, the same $c$ can have different stabilization effects in different brain networks. To ensure consistency across studies, we suggest to make $c$ dependent on $|\lambda(\bm{A})_{max}|$, for instance by $c = 0.01 \cdot |\lambda(\bm{A})_{max}|$. Finally, we observed that the composition and size of the control region set substantially influenced state and control trajectories. The decision critically depends on the individual research question and hence, should be well informed by theoretical or practical considerations. From a methodological perspective, it is important to control a sufficiently large number of brain regions to robustly estimate control energies. In sum, these recommendations could guide more informed modeling choices in future applications of network control theory to pressing questions in cognitive, developmental, and clinical neuroscience.
\\
\\
\textbf{The role of time in network controllability.} In our examination of different modeling choices, we found that both a short time horizon and a fast stabilization of the system induced an alternative control regime. We suggest a common mechanism underlying both time-related observations. Whereas the injected control input has time to diffuse along inter-connections between brain regions over longer time horizons, it might be possible that this diffusion process is constrained over short time horizons. Instead, a different control regime could come into effect in which the injected input primarily controls each brain region independently rather than capitalizing on their interconnections. This finding suggests that time might play a more important role in the controllability of structural brain networks than is commonly assumed. Thus, it could be interesting to further investigate the factor of time, for instance by linking control to real-time measures of brain function \cite{cornblath2018context, stiso2018white}. Another potentially fruitful venture could be to determine optimal control horizons by capitalizing on the natural dynamics of the system or by changing inter-connections in more advanced dynamic models \cite{li2017fundamental}. Such methods emphasizing the role of time could help to develop minimal clinical interventions such as neuromodulation \cite{afshar2011advancing}, which is immediately relevant for the control of seizures in epilepsy \cite{ching2012distributed,stanslaski2011emerging,taylor2015optimal,olmi2019controlling,ehrens2015closed}. The temporal nature of control is also potentially relevant for further refining brain-machine interfaces \cite{gowda2012parameter,stiso2019learning}. 
\\
\\
\textbf{Future directions for proposed model extensions.} Moreover, the present work provides several potentially useful extensions of network control theory. We first developed and validated a complementary measure of structural connectivity motivated by the fact that brain networks based on diffusion imaging data disregard the potential for neural signals to diffuse between spatially adjacent brain regions. We demonstrated that this alternative structural connectivity measure based on the amount of shared neighborhood between two brain regions was complementary to the tractography version. We further showed that their combination introduced more inter-individual variability in controllability metrics, motivating future efforts to employ this approach in studies of individual differences. An important next step is to test whether structural brain networks based on both diffusion imaging and spatial adjacency outperform networks purely based on diffusion imaging data by better accounting for the observed neural dynamics \cite{honey2009predicting, galan2008network}. Additionally, we examined the ability of the brain to control slow and fast dynamics. We found that the capability of a brain region to control different fast modes depended on the specific definition of the control task and was not consistent between time-systems. Neuroscientists interested in the speed of neural changes such as different frequency bands \cite{siegel2012spectral, baillet2017magnetoencephalography} should be careful in justifying their choice of time system and the threshold which defines slow versus fast modes.
\\
\\
Lastly, we wished to extend the existing set of controllability metrics. For this purpose, we developed and validated a new metric that quantifies the complexity of the energy landscape of a given brain network. In other words, the metric measures how heterogeneous all possible state transitions are in the control energy that they require. We showed that the brain exhibited a more homogeneous energy landscape compared to two different null models. We found that both the brain networks' strength distribution and spatial embedding partially explained this observation, which is in line with previous findings connecting local and global network characteristics to network controllability \cite{gu2015controllability, kim2018role, betzel2016optimally}. The requirement of a similar amount of energy to enable diverse state transitions implies that brain architecture supports diverse transitions, which in turn could explain the complex functional dynamics consistently observed in neural systems. A crucial next step is to test the practical utility of this new metric by linking it to development, cognition, and psychiatric disorders. Taken together, the proposed model extensions hopefully stimulate and enrich future research.
\\
\\
\textbf{Expanding horizons of network control theory.} New developments in network control theory are constantly expanding the horizons of research questions that can be tackled with the associated tools. Many (but perhaps not all) of these developments could be helpful in the study of the mind and brain. Efforts have recently revealed a relation between controllability and symmetry \cite{whalen2015observability,zhao2015intrinsic,whalen2016effects}, which could prove useful in determining the impact of bilateral and other symmetries on neural dynamics. The field has begun considering multiobjective functions, tradeoffs, and constraints in control \cite{tang2016robust,keren2014controlling}, in addition to probing a system's potential for control via local topological information \cite{li2018enabling}. As the field of neuroscience moves more concertedly towards multimodal approaches, efforts in the control of multilayer networks could prove particularly useful \cite{menichetti2016control}, as could methods for detecting control nodes across scales \cite{tang2013multiobjective,tang2012constrained,tang2012identifying}. For some questions, advances in the control of nonlinear systems could prove effective \cite{cornelius2013realistic,motter2015networkcontrology,zanudo2017structure}, including applications of Ising models \cite{lynn2017statistical,gu2018energy} and considerations related to the dynamics of neural mass models \cite{liu2017controllability}. Finally, moving beyond network controllability, recent work expanding system identification methods to identify specific form of nonlinear dynamics present in brain is particularly promising \cite{becker2018large,ashourvan2019dynamical,koppe2019identifying}. 
\\
\\
\textbf{Methodological considerations.} Several methodological aspects could potentially constrain the interpretability of our results. First, we capitalized on high-resolution diffusion weighted imaging data for the construction of structural connectivity networks. Associated tractography algorithms are still limited in their capacity to reliably track fiber bundles, particularly long-range connections \cite{thomas2014anatomical, reveley2015superficial}, in terms of their origin, exact direction, and intersection \cite{jbabdi2011tractography}. Nevertheless, diffusion weighted imaging serves as the state-of-the art method to study white matter architecture in humans and therefore, tractography algorithms are continuously being refined \cite{pestilli2014evaluation}. Second, our dynamic model of neural processes relied on several simplifying assumptions including linearity and time-invariance. However, such basic models often provide a good starting-point to approximate higher-order dynamics \cite{honey2009predicting, galan2008network} and can subsequently be adapted to contain more complex features such as non-linearity \cite{fiedler2013dynamics, zanudo2017structure} and time-dependence \cite{li2017fundamental}. Third, it was beyond the scope of this paper to examine the impact of modeling choices in all of their theoretically possible combinations. Instead, we systematically varied one modeling choice at a time while keeping all other choices constant. Thus, the obtained results might not automatically generalize to left-out choices, for example in the presence of higher order interactions. A further limitation of our study is the investigation of control energies in a restricted set of six state transitions. In addition, the initial and target brains states were constructed in a controlled, yet unnatural way by capitalizing on the artificial activation of brain regions belonging to the same cognitive system. For greater ecological validity, future studies could instead rely on real brain states as measured by functional neuroimaging \cite{cornblath2018context, stiso2018white}. Lastly, we wish to point out that the high consistency between minimum and optimal control energy could also be due to different scales of distance and energy costs. To avoid such effects, future efforts could develop an optimal energy algorithm that balances both constraints equally independent of their scale. 
\\
\\
\textbf{Conclusions.} Our systematic overview of network control theory and possible modeling choices aimed to facilitate a deeper understanding and better evaluation of network control theory applications in neuroscience. Future work can potentially benefit from our specific recommendations and the proposed model extensions. Overall, this work hopefully inspires the neuroscience community to fully exploit the potential of network control theory on multiple spatial scales ranging from single neurons to brain regions. Ultimately, such endeavors could advance our understanding of how the architecture of the brain gives rise to complex neural dynamics.
\\

\section{Conflict of Interest}
The authors declare no competing interests.

\section{Acknowledgements}
TMK was supported by the International Research Training Group (IRTG 2150) of the German Research Foundation and by a full doctoral scholarship of the German National Academic foundation. We gratefully acknowledge Eli J. Cornblath, Xiaosong He, Urs Braun, and Lorenzo Caciagli for useful discussions. DSB acknowledges support from the John D. and Catherine T. MacArthur Foundation, the Alfred P. Sloan Foundation, the ISI Foundation, the Paul Allen Foundation, the Army Research Laboratory (W911NF-10-2-0022), the Army Research Office (Bassett-W911NF-14-1-0679, DCIST- W911NF-17-2-0181, Grafton-W911NF-16-1-0474), the Office of Naval Research, the National Institute of Mental Health (2-R01-DC-009209-11, R01-MH112847, R01-MH107235, R21-M MH-106799), the National Institute of Child Health and Human Development (1R01HD086888-01), National Institute of Neurological Disorders and Stroke (R01 NS099348), and the National Science Foundation (BCS-1441502, BCS-1430087, NSF PHY-1554488 and BCS-1631550). The content is solely the responsibility of the authors and does not necessarily represent the official views of any of the funding agencies. 

\section{Author Contributions}
DSB, JS, and TMK designed the study. AEK preprocessed the data. JZK contributed analytic solutions. TMK and JZK wrote code. TMK analyzed the data and wrote the manuscript. All authors edited the manuscript approved the final version.

\newpage
\bibliographystyle{elsarticle-num}
\bibliography{controllability} 

\end{document}